\documentclass[twocolumn,showpacs,preprintnumbers,amsmath,amssymb]{revtex4}

\usepackage{graphicx}
\usepackage{dcolumn}
\usepackage{bm}

\newcommand{\ket}[1]{|#1\rangle}  
  
\newcommand{\brkt}[2]{\langle#1|#2\rangle}  
\newcommand{\eq}[1]{Eq.~(\ref{#1})}  
  
\newcommand{\be}[1]{\begin{equation}\label{#1}}  
\newcommand{\ee}{\end{equation}}  
\begin{document}

\preprint{APS/123-QED}

\title{Femtosecond Photodissociation of Molecules Facilitated by Noise}

\author{Kamal P. Singh, Anatole Kenfack and Jan M.~Rost}
 \affiliation{ Max Planck Institute for the Physics of Complex Systems, N\"othnitzer Strasse 38, 01187 Dresden, Germany.}

\date{\today}

\begin{abstract}
We investigate the dynamics of diatomic molecules
 subjected to both a femtosecond mid-infrared laser pulse and
 Gaussian white noise.  The stochastic Schr\"odinger equation with a
 Morse potential is used to describe the molecular vibrations under noise
 and the laser pulse.  For weak laser intensity, well below the
 dissociation threshold, it is shown that one can find an optimum
 amount of noise that leads to a dramatic enhancement of the
 dissociation probability.  The enhancement landscape which is shown
 as a function of both the noise and the laser strength, exhibits a
 global maximum.  A frequency-resolved gain profile is recorded
 with a pump-probe set-up which is experimentally realizable.
 With this profile we identify the linear and nonlinear multiphoton
 processes created by the interplay between laser and noise and assess
 their relative contribution to the dissociation enhancement.
\end{abstract}

\pacs{02.50.Ey, 33.80.Gj, 42.50.Hz}
\maketitle
\section{Introduction}

 The efficient dissociation of a molecular bond by the application of ultrashort
 laser pulses is a long sought after goal in photochemistry \cite{review, Zewail}.
 However, to achieve significant molecular dissociation with coherent 
 ultrashort laser pulses is difficult due to the
 anharmonicity of the molecular vibrations \cite{bloembergen}.
 Very high intensities of the laser field are therefore required 
 to produce  significant dissociation, despite the lower value of
 the dissociation energy compared to the ionization potential. 
 For such strong laser fields molecular tunnel ionization 
 often masks the dissociation process \cite{kim, TunlIonz}. 

 Alternative strategies have been proposed  to enhance the dissociation yield for
 weaker laser pulses. For instance, the approach using the
 linearly (or circularly) polarized
 chirped laser pulses have been proposed. Here, the frequency is
 designed to match the vibrational ladder of a specific molecule,
 thereby facilitating the ladder climbing of the molecule and
 eventually its dissociation \cite{cheal90}.
 These ideas have been confirmed in the experiments, 
 using chirped femtosecond mid-infrared (MIR) laser pulses. A large population
 transfer to higher vibrational levels have been achieved with diatomic molecules
 ($HCl$, $NO$ etc) as well as in some polyatomic molecules
 ($Cr{(CO)_6}$) \cite{Corkum, Zewail2, Kompa2}. 
 Furthermore, the optimally tailored pulses by a
 closed loop control have also been used quite recently to dissociate a
 specific molecular bond in a polyatomic molecule; it was possible to control in particular
 the branching ratio between two possible reaction channels \cite{Baumart}. 

 Following a completely different approach to achieve the molecular dissociation,
 it has been proposed recently that one can instead use an incoherent 
 random kicks \cite{kenny}. In the same spirit, there has been a growing 
 interest in understanding the role of
 random fluctuations on the quantum dynamics of atomic and molecular
 systems \cite{Rabitz1, singhkp, Rabitz2}. 
 However, the molecular dissociation processes due to the simultaneous
 application of a laser and a noise source has so far not yet
 been considered. In doing so, we expect the coherent laser pulse to
 become more efficient when submerged in a small amount of noise. It
 is indeed well known that in some classical as well as quantum nonlinear systems
 (e.g. bistable systems), the noise can lower the nonlinear threshold via the
 stochastic resonance (SR) effect \cite{Gamat, Buchl}. It is also worth mentioning
 here that SR also exists in some nonlinear chemical reactions \cite{Gamat,SRchem}, 
 but has not been demonstrated in the crucial process of molecular (photo)dissociation
 on a femtosecond time-scale yet.

 Here, we propose an approach to control the molecular dissociation by irradiating
 diatomic molecules by a femtosecond MIR laser pulse in the presence of noise.
 We show how and to what extent the presence of noise can reduce the dissociation
 threshold of diatomic molecules due to its nonlinear interaction 
 with the laser pulse. This approach can be applicable to both polar and nonpolar 
 molecules and is readily possible to use in experiments.  

 We begin the main text of this work in Sec. II with a brief description of the
 molecular Hamiltonian including a coherent laser pulse and a stochastic term. 
 In Sec. III we describe the fast-Fourier transform (FFT) split-operator method to
 solve the stochastic Schrodinger equation with absorbing boundary conditions, 
 and focus on the relevant physical observable of interest. 
 Our results for the role of noise in the molecular photodissociation, and 
 other interesting features of the effect are discussed in Sec. IV.    
 Finally, Sec. V concludes the paper with some perspectives.

\section{The theoretical approach}  

The total molecular Hamiltonian for the diatomic molecule under
time-dependent external fields is given by (atomic units are used
unless stated otherwise)
\begin{equation}  
 H(t)=H_{0}-x\xi(t) - xF(t),  
\label{hamiltonian}  
\end{equation}  
 where $H_{0}=\frac{p^{2}}{2m}+V(x)$ describes vibrational motion of 
 the molecule with reduced mass $m$ in the Morse potential \cite{morse}  
\begin{equation}  
V(x)=-D_{e}+D_{e}\left( 1-\exp(-\beta x)\right)^{2},  
\label{potential}  
\end{equation}  
with  well depth $D_e$ and length scale $\beta$.   
The eigenenergies $E_n$ of the Morse oscillator $H_0$ are given by  
\be{eigenvalues}  
E_n=\hbar\omega_e(n+1/2)[1-B(n+1/2)/2],\,\,\,\,\,\,\,\,\,0\le n\le [j]  
\end{equation}  
where $\omega_e$ is the harmonic frequency, $n_b = [j]+1$ is the
number of bound states with \be{parameters} j=1/B-1/2,\,\,\,B =
\hbar\beta(2mD_e)^{-1/2},\,\,\,\hbar\omega_e=2BD_e\,.
\end{equation}  

\begin{figure}[t]
\includegraphics[width=.95\columnwidth]{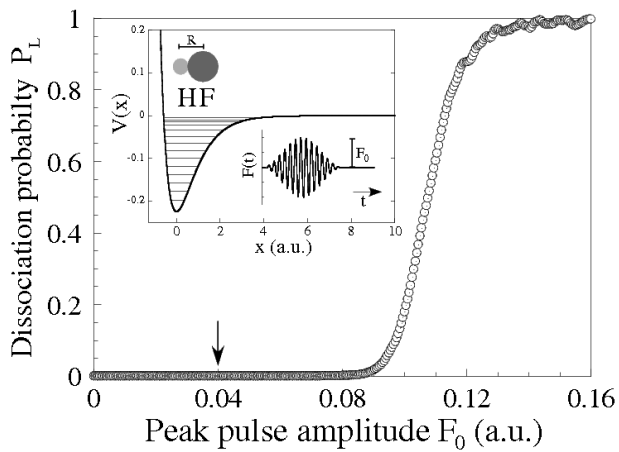}
 \caption{
 The dissociation probability $P_L$ as function of the laser peak 
 amplitude $F_0$ over long time t ($t>>T_p$). Here, $\omega=0.007$~a.u.,
 $\delta=0.0$, and $T_p=30\pi/\omega$. The arrow indicates the value of
 $F_0=0.04$~a.u. for which $P_L\sim 0$ is far less than the dissociation
 threshold. The inset shows the corresponding $15$ optical cycle laser pulse
 $F(t)$ (light solid line) and the schematic pictures of a diatomic
 molecule $HF$ (two nuclei separated by $R$, with $R_0$, being the
 equilibrium internuclear distance) and the Morse potential $V(x)$
 (thick solid line), with $x=R-R_0$. }
\end{figure}

 The laser field is a nonresonant MIR femtosecond pulse, 
 \begin{equation}  
 \mbox{$ F(t) = f(t) F_0\sin(\omega t+\delta)$}\,,
 \label{Lsrpulse}
 \end{equation}  
 where  $\omega$ denotes
 the angular frequency, and the effective peak amplitude $F_0$ 
 as well as the noise amplitude $\xi(t)$
 (see below) contain the molecular dipole gradient \cite{kenny}.
 We choose a smooth pulse envelope $f(t)$ of the form
 \begin{equation}  
 \mbox{$ f(t) = \sin^{2}(\pi t/ T_p)$},
 \label{envelop}
 \end{equation}  
 where $T_p$ is the pulse duration of typically 15 optical cycles (see
 inset of Fig.~1).

 The noise term $\xi(t)$ is a zero-mean
 Gaussian white noise having the following properties,
 \begin{equation}   
   \langle \xi(t) \rangle = 0,
   \label{eqn:eqn2a}
 \end{equation}    
 \begin{equation}   
   \langle \xi(t)\xi(t')\rangle = 2D\;\delta (t-t'),
   \label{eqn:eqn2b}
 \end{equation}                  
 where $D$ is the noise intensity \cite{book}.

\section{Time evolution and Observables}  

\subsection{Quantum stochastic propagation}  

The time-evolution under \eq{hamiltonian} is non-deterministic due to
the stochastic nature of the Hamiltonian.  A  solution taking into 
account the external fields perturbatively is not possible, because 
the fields of interest are so strong that
they
influence the dynamics substantially.  Hence, we have to solve the
full time-dependent stochastic Schr\"odinger equation,
\begin{eqnarray}
 i\hbar\frac{\partial \ket{\psi(t)}}{\partial
 t}=H(t)\ket{\psi(t)}\,\label{q_langevin}.
 \end{eqnarray} 
The solution is accomplished by averaging over a sufficient number of
deterministic solutions under different specific realizations $r$ of
the noise.

For a given realization $r$, the solution of the stochastic
Schr\"odinger equation amounts to solve the standard time-dependent
Schr\"odinger equation,
 \begin{eqnarray} 
     \ket{\psi_{r}(t)}=U_{r}(t,t_{0})\ket{\psi(t_{0})}, \label{e36} 
 \end{eqnarray}
starting from an initial state $|\psi(t_0)\rangle$  at time $t_{0}$.
The stochastic  evolution operator $U_{r}(t,t_{0})$ is written as a 
product of operators propagating over a  small time interval $\Delta 
t$,
\begin{eqnarray} 
 \mbox{$U_r (\Delta t) = \exp\left( -i\int_{t}^{t+\Delta t} H(x,t) dt\right)$}.
 \label{Spropagtr1}
\end{eqnarray}  
This short-time propagator can be formulated explicitely by 
evaluating the integral in the exponent in the
Stratonovitch sense \cite{book}, leading to
\begin{equation} 
  U_r(\Delta t)=\exp( -i(H_0(x)-xF(t))\Delta t) \exp( i x\sqrt{2D\Delta t}\xi_t),
  \label{Spropagtr2}
 \end{equation} 
where $\xi_t$ is a Gaussian distributed random number of unit
variance.  The representation of \eq{Spropagtr2} illustrates how the
coherent evolution of the system (with laser) aquires an additional
random phase in the form of a momentum kick whose strength is random due
to $\xi_{t}$ but scales with the amplitude of the noise
$\sqrt{2D\Delta t}$ over the time interval $\Delta t$.  The short-time propagator
given by \eq{Spropagtr2} can be easily realized
numerically using the FFT split-operator approach with absorbing
boundary conditions \cite{feit}.

\subsection{Dissociation probability}  
We use the dissociation probability as a measure for the non-linear
coupling of energy from the external fields to the molecule.  It can
be most conveniently expressed as a complement of the probability to find the molecule
at time $t$ in any of its bound states, which reads for a given
realization $r$

 \begin{eqnarray} P_{d}^r(t)=1- \sum_{\nu=0}^{N_b-1}\left|\brkt{\psi_\nu}{\psi_r(t)}\right|^{2}\,. 
 \label{e314} \end{eqnarray}  
 where for the initial state $\ket{\psi_0(t=0)}$ we take the ground state
 of the Morse potential with energy $E_0$ (Eq.~3). 
  Typically we will average over a large number of realization ($N_r$ about 100) to calculate
 the average dissociation probability over long time $t_\infty>>T_p$,
  \begin{eqnarray}  
  P = \frac{1}{N_r} \sum_{r=0}^{N_r} P_{d}^{r}(t_\infty)\,. \label{ePD} 
  \end{eqnarray} 
  In the following, we shall consider different physical situations
  such as when the molecule is subjected to (i) the laser pulse alone
  (denoted by subsrcript L), (ii) the noise alone (N), or (iii) a
  combination of both fields (L+N).

\section{Results and discussions}  
\subsection{Femtosecond photodissociation under noise}

  Photodissociation of diatomic molecules by a femtosecond laser pulse
  is a highly non-linear processes, as can be seen in Fig.~1 for a 15
  cycle MIR pulse at $\omega=0.007$~a.u..  The dissociation probability
  $P_{L}$ versus the peak field amplitude $F_0$ exhibits a
  prototypical S-shape curve with a threshold like behavior. High
  intensities (i.e., $F_{0}^{2}$ values) are necessary for dissociation by a laser
  pulse alone. However, the addition of a weak amplitude noise to 
 a subthreshold laser pulse may lead to substantial dissociation.
 As we have shown before, such a combination is quite efficient for 
 an otpimum choice of the noise amplitude in achieving ionization of 
 an atom \cite{singhkp}, resembling the phenomenon of stochastic resonance
 \cite{Gamat,Buchl}.

 \begin{figure}[t]
\includegraphics[width=.95\columnwidth]{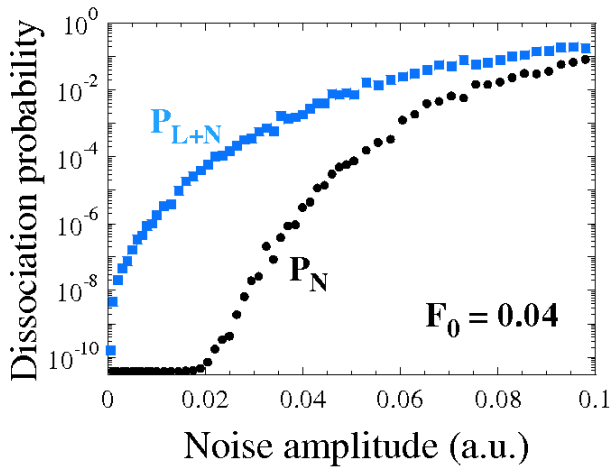}
 \caption{ The average dissociation probability of the molecule for
 noise only $P_N$ (square) and for the simultaneous application of the
 laser pulse and the noise $P_{L+N}$ (circle).  The total number of
 realization $N_r=100$, and the laser peak amplitude is kept fixed at
 $F_0=0.04$~a.u..}
\end{figure}

 We choose a laser pulse of peak amplitude $F_0=0.04$~a.u., for which the
 dissociation probability is extremely small, $P_{L} <10^{-10}$ (see
 Fig.~1).  Keeping this laser pulse unchanged, we add a small amount
 of noise to it.  Fig.~2 shows the average dissociation probability
 $P_{L+N}$ versus the noise amplitude $\sqrt D$ (squares).  One can
 clearly see that by adding a small amount of noise, $P_{L+N}$ rises
 rapidly, well beyond the increase induced by the noise alone ($P_{N}$,
 circles) which is applied for the same time $T_{p}$ as the laser
 pulse.

 The enhancement by noise can be quantified with the parameter
 \begin{equation}            
 \eta = \frac{P_{L+N} - P_0}{P_0} 
 \label{eqn:eqn4}
 \end{equation}
  with $P_{0} = P_L + P_N $
  which is plotted in Fig.~3. By construction, $\eta\to 0$ if one of 
  the two fields, laser or noise, is much larger than the other.
  However, for a specfic noise amplitude, here at $\sqrt D = 0.02$, a 
  dramatic enhancement of the order of $10^{5}$
 is  observed. 

 \begin{figure}[t]   
\includegraphics[width=.95\columnwidth]{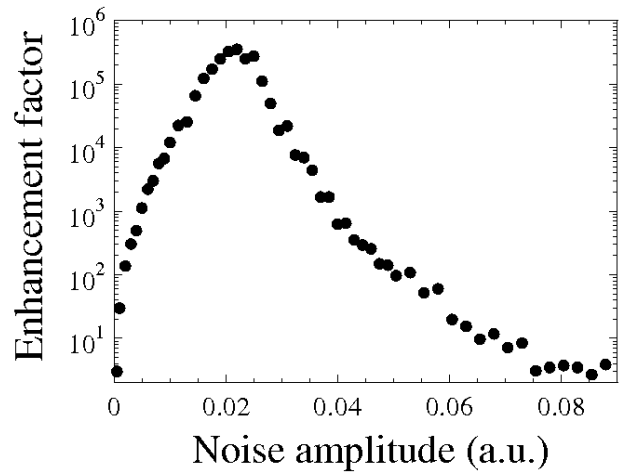}
 \caption{
 The factor of enhancement $\eta$ versus the noise amplitude $\sqrt D$ for the 15
 cycle long MIR pulse with $F_0=0.04$~a.u. and $\omega=0.007$~a.u..
 }
\end{figure}
\subsection{The net enhancement landscape} 
 Having kept the laser amplitude fixed so far to a subthreshold value
 ($F_{0}=0.04$~a.u.), we now investigate the enhancement factor as a
 function of both, laser and noise amplitudes.  As can be seen in
 Fig.~4, $\eta(\sqrt D, F_0)$ shows a global maximum.  Additional
 calculations have revealed that the shape of $\eta(\sqrt D, F_0)$ is
 quite robust for different range of parameters, such as
 carrier-envelop phase $\delta$, pulse duration $T_p$, and even for
 different values of the laser frequency $\omega$ in the infrared (IR)
 regime.  Hence, the enhancement landscape from Fig.~4 can be used as
 a guide to identify the range of enhancement possible in experiments
 on molecular photodissociation under noise.
 \begin{figure}[t]   
\includegraphics[width=.95\columnwidth]{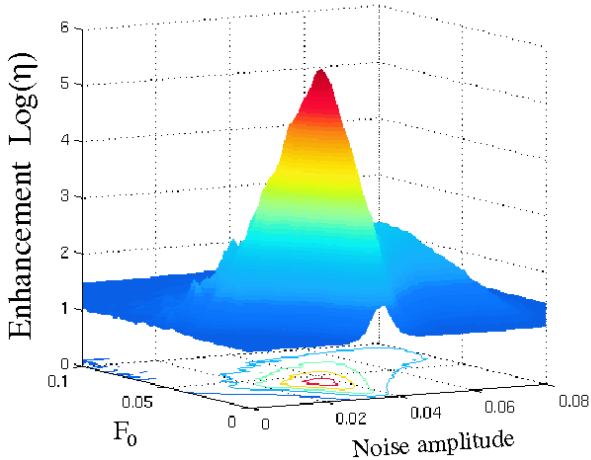}
 \caption{ The enhancement landscape $\eta(\sqrt D, F_0)$ versus the
 peak pulse amplitude $F_0$ and the noise amplitude $\sqrt D$.  Other
 parameters are kept fixed as in figure 3.  Contour lines of equal
 enhancement are also shown on the ($\sqrt D$,$F_0$) plane.}
 \end{figure}

\subsection{Frequency-resolved enhancement profile for $HF$} 
 \begin{figure}[t]   
\includegraphics[width=.95\columnwidth]{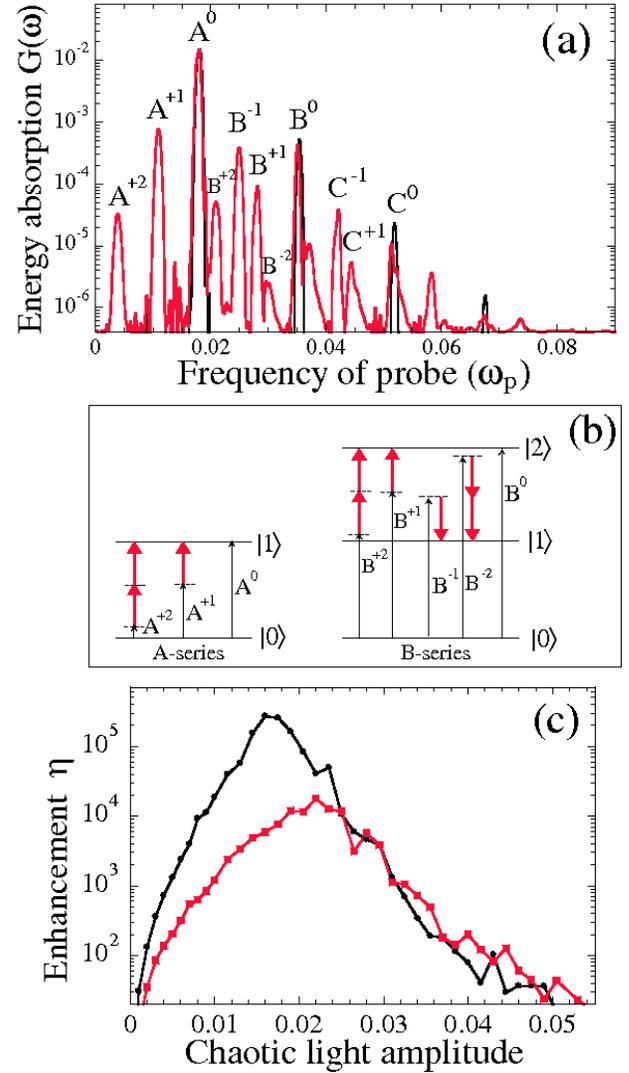}
 \caption{ (a) The frequency-resolved gain $G(\omega)$ for the driven
 $HF$ molecule measured using a simultaneous pump-probe setting
 without noise.  $G(\omega)$ for the bare molecule is also shown
 (black curve) for comparison.  (b) Various single-photon and
 mutiphoton transitions corresponding to the peaks in $G(\omega)$ are
 classified as A-series, B-series, C-series that involve first two,
 three, four vibrational levels of $HF$, respectively.  Thick red
 arrow denote the pump photon ($\omega=0.007$~a.u.) and thin black arrow is
 for tunable weak probe pulse.  Pulse bandwidth is here negligible
 compared to the photon energies.  (c) Enhancement curve for the
 chaotic light with perforated spectrum of hole-widths 0.004~a.u. is
 compared with the broadband light of BW=1.0~a.u..  }
\end{figure}
In order to shed light on the mechanism underlying the stochastic
enhancement of dissociation we take a closer look at the frequency
resolved enhancement profile for the case of $HF$.  This will reveal
which frequencies provide the gain within the braod band noise and
will in turn clarify the non-linear photo processes responsible for
the enhancement.

For this purpose, we record the frequency-resolved molecular gain
$G(\omega)$ (FRMG) employing a pump-probe type of setting \emph{without}
noise.  We consider the $HF$ molecule to be driven by the MIR
laser pulse of $F_0=0.04$~a.u. and $\omega=0.007$~a.u. (pump pulse), and replace
the noise by a probe pulse of amplitude $F_p$ and of tunable frequency
$\omega_p$.  The tunable probe pulse has the same duration and
envelope as the fixed pump pulse.  The probe amplitude
($F_p/F_0\simeq 0.05$) is such that it can only produce single photon
transitions.  For a $HF$ molecule in its ground state, $G(\omega)$ is
obtained by measuring the net energy absorbed as a function of
$\omega_p$ and is shown in Fig.~5(a).  It exhibits several clearly
resolved peaks which play a dominant role in the dissociation enhancement  
under broad band light.

 We have identified these peaks as originating from single or multiple
 photon processes for the level-structure of $HF$ molecule.  For
 convenience, these processes can be classified as follows.  All
 photo processes that involve the lowest two levels of the $HF$ are
 termed A-series, while the ones involving the next higher level
 (three levels) are termed B-series, and so on.  Fig.  5(b) shows
 these excitation pathways for the given pump frequency
 $\omega=0.007$~a.u..  These basically involve n-photons of the pump
 ($n=0,\pm1,\pm2,..$) plus a single photon of the probe.  Note that in
 addition to the resonant single photon transitions, second- and
 third-order processes play an important role.  Thus, the diatomic
 molecule offers an interesting example where the role of higher order
 photo-processes can be clearly isolated, which is nontrivial in the
 case of an analogous atomic system \cite{singhkp}.

 To highlight the contribution of the multiphoton transitions in the
 enhancement, we have eliminated the first four single-photon resonant
 frequencies from the noise spectrum.  Such a filtered noise can be
 easily simulated by means of the chaotic light \cite{singhkp}.  For a
 chosen hole-width of $0.008$~a.u. centered at the resonant
 frequencies of $HF$, a drop of more than one order of magnitude in
 the maximum $\eta$ is observed as shown in Fig.  5(c).  Note that one
 is still left with a significant enhancement ($\eta \simeq 10^4$),
 which can be clearly attributed to the higher order processes other
 than single photon resonant transitions.  Although we have considered
 specifically $HF$, the approach used here can be applied to other
 diatomic molecules.  We have found similar enhancement features for
 $HCl$ and $H_2$ molecules within the framework of the Morse potential
 subjected to different MIR pulses.

\section{Summary and Conclusion}  
 We have shown that it is possible to significantly enhance the
 dissociation probability of a diatomic molecule under a weak
 femtosecond MIR pulse by adding a small amount of noise to it.  The
 net enhancement for a given laser pulse exhibits a single maximum as
 the noise level is varied.  This maximum suggests an optimum noise
 amplitude to achieve the maximum dissociation, in analogy to the
 stochastic resonance phenomenon.  The enhancement landscape, that is
 the enhancement as a function of laser amplitude and noise amplitude,
 exhibits a global maximum.  Analyzing the frequency resolved gain
 profile of the molecule using a pump-probe setting we have identified
 the dominant frequencies and the corresponding physical processes
 activated by the noise.  We conclude that in addition to
 single-photon resonant transitions, multiphoton transitions play a
 significant role.

 Similar effects are expected for other diatomic molecules if exposed
 to a combination of noise and laser pulse,
 where pulse duration and wavelength in the MIR regime may be chosen
 from a wide parameter range.  Of particular interest in the future
 will be if the proposed nonlinear interplay can achieve selective
 bond breaking by enhancing dissociation of a particular
 bond in a polyatomic molecule.

\begin{acknowledgments}  
We gratefully acknowledge fruitful discussions with A. D. Bandrauk and N. Singh.
AK would like to acknowledge the financial supports by the Max-Planck 
Gessellschaft through Reimar L\"ust fund and by the Alexander von Humboldt
 foundation with the research grant No.IV. 4-KAM 1068533 STP.
\end{acknowledgments}

\end{document}